\documentclass[conference,10pt]{IEEEtran}
\IEEEoverridecommandlockouts
\usepackage{cite}
\usepackage{amsmath,amssymb,amsfonts}
\usepackage{algorithmic}
\usepackage{graphicx,physics,array}
\usepackage{textcomp}
\usepackage{xcolor,soul}
\usepackage{multirow}
\usepackage{hyperref}
\usepackage{adjustbox}
\usepackage{tikz}
\newcommand*\circled[1]{\tikz[baseline=(char.base)]{
            \node[shape=circle,fill,inner sep=0.5pt] (char) {\textcolor{white}{#1}};}}
\def\BibTeX{{\rm B\kern-.05em{\sc i\kern-.025em b}\kern-.08em
    T\kern-.1667em\lower.7ex\hbox{E}\kern-.125emX}}

\begin{document}

\title{Magic Mirror on the Wall, How to Benchmark Quantum Error Correction Codes, Overall ?
}

\author{\IEEEauthorblockN{Avimita Chatterjee}
\IEEEauthorblockA{\textit{Department of Computer Science \& Engineering} \\
\textit{Pennsylvania State University}\\
PA, USA\\
amc8313@psu.edu}
\and
\IEEEauthorblockN{Swaroop Ghosh}
\IEEEauthorblockA{\textit{School of EECS} \\
\textit{Pennsylvania State University}\\
PA, USA\\
szg212@psu.edu}
}

\maketitle

\begin{abstract}
Quantum Error Correction Codes (QECCs) are pivotal in advancing quantum computing by protecting quantum states against the adverse effects of noise and errors. With a variety of QECCs developed, including new developments and modifications of existing ones, selecting an appropriate QECC tailored to specific conditions is crucial. Despite significant improvements in the field of QECCs, a unified methodology for evaluating them consistently has remained elusive. This paper addresses this gap by introducing a set of eight universal parameters and evaluating nine prominent QECCs for these parameters. We establish a universal benchmarking approach %\hl{incorporating diverse QECCs including the cutting-edge Bivariate Bicycle (BB) codes recently unveiled by IBM}, 
and highlight the complexity of quantum error correction, indicating that the choice of a QECC depends on each scenario's unique requirements and limitations.
Furthermore, we develop a systematic strategy for selecting QECCs that adapts to the specific requirements of a given scenario, facilitating a tailored approach to quantum error correction. Additionally, we introduce a QECC recommendation tool that assesses the characteristics of a given scenario provided by the user, subsequently recommending a spectrum of QECCs from most to least suitable, along with the maximum achievable distance for each code. This tool is designed to be adaptable, allowing for the inclusion of new QECCs and the modification of their parameters with minimal effort, ensuring its relevance in the evolving landscape of quantum computing.
\end{abstract}


\section{Introduction} \label{intro}

Quantum computing harnesses the principles of quantum mechanics to execute tasks that are beyond the capabilities of classical computing. This advanced technology finds applications in various fields, including molecule simulation for drug development, financial optimization enhancements, machine learning acceleration, improvement of optimization tasks, and revolutionary changes in supply chain management, among others \cite{reiher2017elucidating, orus2019quantum, schuld2015introduction, singh2020nebula, gachnang2022quantum, ajagekar2019quantum}. However, commercialization of these quantum computing developments incurs challenges such as qubit stability and quantum noise \cite{clerk2010introduction, mouloudakis2021entanglement}. Quantum Error Correction Codes (QECCs) are crucial for achieving fault-tolerant quantum computing in the face of unavoidable qubit noise \cite{lo1998introduction, devitt2013quantum}. Traditional error correction methods \cite{hamming1950error} encounter unique challenges in the quantum realm, primarily due to constraints like the no-cloning theorem \cite{wootters2009no} and the wave-function collapse that occurs during qubit measurement \cite{von2018mathematical}. Research in this field has led to the creation of various quantum codes, such as the five-qubit, Bacon-Shor, topological, surface, color, and heavy-hexagon code \cite{sundaresan2022matching, tumpa2023federated, abobeih2022fault, bacon2006operator, kitaev1997quantum, krinner2022realizing, bombin2006topological}.

\subsection{Motivation}

The landscape of quantum computing is rich with an array of QECCs, each heralding its unique set of advantages and compromises. From the foundational Shor and Steane codes to the cutting-edge realms of topological codes, the diversity of available QECCs presents a pivotal question: ``Which code is most suitable for a specific quantum computing scenario?" This query is at the heart of advancing quantum computing technologies and underscores the pressing need for a comprehensive and standardized benchmarking methodology. Despite the wealth of research assessing various QECCs, the absence of a unified framework for objective and comparative analysis across different codes stands as a significant gap. Such comparative insight is indispensable not only for the ongoing research within the quantum computing sphere but also for steering the direction of future QECC development.

Navigating through the myriad studies to deduce the optimal QECC for a particular scenario poses a formidable challenge, compounded further when multiple scenarios are considered. Let us consider two hypothetical scenarios: (a) we have a setup with superconducting hardware with $100$ qubits where the circuit requiring correction involves just a single qubit and (b) we have a setup with trapped-ion hardware with $600$ qubits, and circuit to be corrected includes $5$ qubits as well as multi-qubit gates. The optimal QECC for these two situations would differ significantly. The complexity of determining the most appropriate QECC — factoring in the types of qubits, error rates, and other critical parameters — demands a systematic selection strategy. The necessity for a framework that can swiftly identify a suitable QECC based on specific scenario requirements is not just a matter of convenience but a crucial step forward in the practical implementation of QECCs.

\subsection{Contribution}

To our knowledge, this research marks the inaugural effort to establish a unified benchmarking methodology for QECCs. Our objective is to discern how different QECCs are ideally suited for various scenarios and devise a systematic strategy for selecting QECCs appropriately. 
%We aim to provide a comprehensive framework to facilitate this process. 
The contributions of this study can be outlined as follows:
\circled{1}
\textit{Comprehensive Analysis of QECCs:} We conduct an in-depth analysis of nine leading QECCs utilizing eight distinct parameters: qubit overhead, threshold, type of error protection, decoding complexity, types of transversal gates, scalability, the range of qubit types it has been realized on, and overall complexity. This enables a detailed comparative overview across different codes.
\circled{2}
\textit{Scenario-dependent QECC Suitability:} Our findings emphasize that the selection of the most suitable QECC is contingent upon the specific scenario of its intended application, rather than a theoretical hierarchy of QECC efficacy. We demonstrate this through the use of radar charts, showcasing the trade-offs involved in choosing different QECCs, including variations within a single QECC type.
\circled{3}
\textit{Universal QECC Selection Strategy:} We introduce a universal, systematic approach for selecting a QECC tailored to a given scenario. This strategy is designed to guide users in identifying the QECC that best matches their specific requirements, considering the multifaceted nature of quantum computing scenarios.
\circled{4}
\textit{An Adaptive QECC Recommendation Tool:} We develop a dynamic tool that assesses user-provided scenario parameters and recommends a list of QECCs, ranked from most to least suitable, along with the maximum distance achievable for each code. This tool can allow for the inclusion of additional QECCs and the swift modification of their parameters without necessitating significant overhauls to the system. This ensures the framework's adaptability and relevance in the rapidly evolving quantum computing landscape. For instance, should a QECC that has not been previously realized on trapped-ion qubits later achieve this milestone, the framework can easily accommodate this new information.

The tool, along with its code base, is available in a public GitHub repository. \footnote{GitHub Repository Link: \href{https://github.com/Avimita-amc8313/Magic-Mirror-on-the-Wall-How-to-Benchmark-Quantum-Error-Correction-Codes-Overall-}{Magic Mirror on the Wall, How to Benchmark Quantum Error Correction Codes, Overall ?}}

\subsection{Paper Structure}

Section \ref{theory} provides a brief introduction to QECCs and an extensive examination of the nine QECCs featured in this research. Section \ref{benchmarking_methods} details the eight benchmarking parameters, selection criteria for codes and parameters, and the comparative analysis method. Section \ref{benchmarking_analysis} offers a thorough analysis of the nine QECCs against the selected parameters. Section \ref{strategy} discusses a simplified code selection strategy and introduces our QECC recommendation tool. The paper concludes in Section \ref{conclusion}.
\section{Theoretical Background} \label{theory}

% This section gives a general overview of the quantum error correction codes and then moves on to describe the codes that have been considered in this study.

\subsection{Brief Overview of QECCs}

QECCs %are pivotal in quantum computing for protecting quantum information from errors due to quantum noise \cite{shor1995scheme}. They 
differ fundamentally from classical error correction as they manage information encoded in quantum states, characterized by properties like superposition and entanglement \cite{nielsen2001quantum}. Quantum systems are notably sensitive, with quantum states being easily disrupted by external disturbances \cite{preskill1998reliable}. This fragility presents a significant hurdle for reliable quantum computation and information storage, effectively addressed by QECCs \cite{gottesman1997stabilizer}. QECCs function by encoding quantum information across multiple qubits, allowing the system to detect and correct errors without directly measuring the quantum state, in compliance with the no-cloning theorem \cite{wootters1982single, dieks1982communication}. Quantum errors primarily include bit-flip and phase-flip errors, with more complex errors combining these two \cite{shor1995scheme}. QECCs correct these errors through entangled states and collective measurements \cite{calderbank1996good}. Various QECCs have been proposed: Shor Code, the first QECC capable of correcting arbitrary single-qubit errors \cite{shor1995scheme}; Steane Code, a Calderbank-Shor-Steane (CSS) code illustrating fault-tolerant quantum computation principles \cite{steane1996error}; Topological Codes, like Kitaev's Toric code, using a lattice of qubits for robustness against local errors \cite{kitaev2003fault} and Surface Codes, known for simpler error correction procedures, suitable for large-scale quantum computations \cite{fowler2012surface}. QECCs are integral to practical quantum computing, particularly in achieving fault-tolerant quantum computation, a vital step for scalable and reliable quantum computers \cite{aharonov1997fault}. Ongoing research in QECCs focuses on optimizing code efficiency, reducing qubit overhead, and enhancing fault tolerance, crucial for the evolution of quantum computing \cite{terhal2015quantum}.

\subsection{Descriptions of QECCs Under Study}

\begin{figure}
    \centering
    \includegraphics[width=0.75\linewidth]{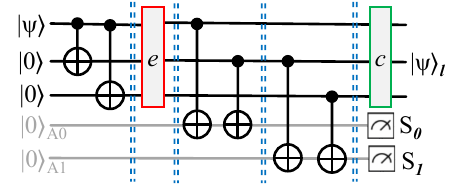}
    \caption{\textbf{Quantum circuit implementing a 3-qubit repetition code.} 
    % It features a state preparation circuit, the possibility of a single bit-flip error on any of the three qubits, two ancilla qubits initialized to $\ket{0}$ for parity checking, a first stabilizer circuit ($Z \otimes Z \otimes I$) for measuring parity between the first two qubits, a second stabilizer circuit ($I \otimes Z \otimes Z$) for the last two qubits, and syndrome bits $S_0, S_1$ obtained via the ancilla qubits for error detection and deduction. Correction is achieved by applying a sequence of self-inverse Pauli-gates ($c$) to the qubit requiring correction.
    }
    \label{fig:rep_code}
    \vspace{-10pt}
\end{figure}

\textbf{Repetition Code:}
The Quantum Repetition Code \cite{devitt2013quantum, peres1985reversible} is one of the simplest forms of QECCs. Its primary function is to protect quantum information against errors, particularly bit-flip errors. This code extends the concept of the classical repetition code into the quantum realm, leveraging the fundamental principles of quantum mechanics \cite{chatterjee2023quantum}. The quantum repetition code works by encoding a single qubit of information across multiple qubits. For instance, in a 3-qubit repetition code, a single logical qubit $\ket{0}$ or $\ket{1}$ is encoded as $\ket{000}$ or $\ket{111}$, respectively. This redundancy allows the code to detect and correct bit-flip errors that might occur on individual qubits. Error detection in the quantum repetition code is accomplished through a process known as syndrome measurement. This involves measuring the parity of neighboring qubits without collapsing the overall quantum state. For example, if the encoded state is $\ket{000}$ and a bit-flip error occurs on the second qubit, the state changes to $\ket{010}$. The syndrome measurement detects this error by noticing the changed parity between the first and second qubits and the second and third qubits. Once detected, quantum operations can be applied to correct the error and restore the original state. Fig. \ref{fig:rep_code} shows a quantum circuit for a 3-qubit repetition code, featuring state preparation, potential single bit-flip error detection, parity checking with two ancilla qubits, and error correction using self-inverse Pauli-gates.

\begin{figure}
    \centering
    \includegraphics[width=0.75\linewidth]{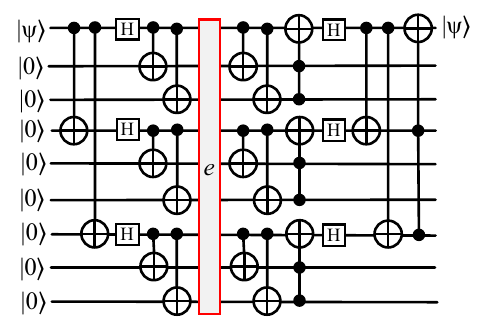}
    \caption{\textbf{Quantum circuit implementing a 9-qubit Shor code.} 
    % It features a state preparation encoding circuit, the possibility of a single bit-flip or phase flip error on any of the three qubits, error correction, and error recovery.
    }
    \label{fig:shor_code}
    \vspace{-10pt}
\end{figure}

\textbf{Shor Code:}
The Shor Code \cite{shor1995scheme} marks a significant advancement in quantum error correction by being the first to correct arbitrary quantum errors, including both bit-flip and phase-flip errors, in a qubit. It encodes one qubit of information into nine physical qubits through a two-layer structure: the first layer addresses bit-flip errors by tripling the qubit, and the second layer corrects phase-flip errors by further encoding each of these into three qubits. This innovative approach allows the Shor Code to correct any single-qubit error across its nine qubits. It is described as a degenerate code, where different error types can affect the codewords similarly. The two basis states of the code are represented as $\ket{0_L}$ and $\ket{1_L}$, intricately constructed from the superposition and entanglement of the nine qubits. The two basis states for the code are defined as: $\ket{0_L} = \frac{1}{\sqrt{8}} (\ket{000} + \ket{111}) (\ket{000} + \ket{111}) (\ket{000} + \ket{111})$ and $\ket{1_L} = \frac{1}{\sqrt{8}} (\ket{000} - \ket{111}) (\ket{000} - \ket{111}) (\ket{000} - \ket{111})$. This structure ensures that the Shor Code corrects one X error per trio of qubits and a single Z error across nine qubits, making it a comprehensive quantum error-correcting code. X errors are managed by a correction circuit for each trio, identifying and fixing bit-flips. Z errors are addressed by comparing sign differences between trios with CNOT gates. Its degeneracy allows it to correct a phase flip in any qubit of a trio effectively. Although capable of correcting three bit flips in separate trios, it's primarily a single-error corrector due to its limitations with multiple errors. Fig. \ref{fig:shor_code} shows the Shor Code's encoding, error introduction, and correction process.

\begin{figure}
    \centering
    \includegraphics[width=0.75\linewidth]{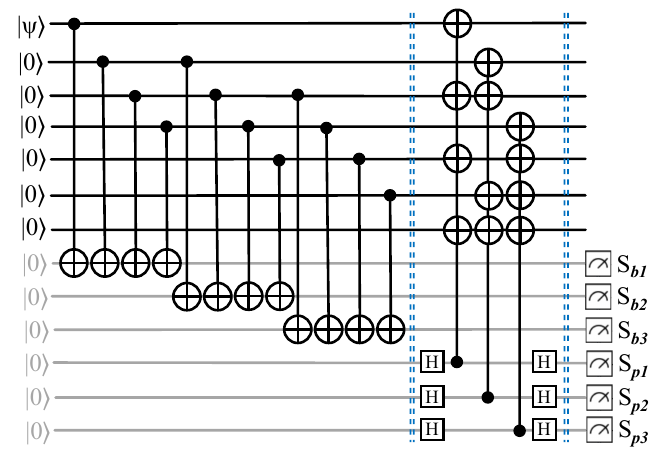}
    \caption{\textbf{Implementation of a 7-qubit Steane code decoding circuit.} 
    % It features three bit-flip detection stabilizers and three phase-flip detection stabilizers along with six ancilla qubits and their respective syndrome measurements (3 for detecting bit-flips and 3 for detecting phase-flips.)
    }
    \label{fig:steane_code}
    \vspace{-10pt}
\end{figure}

\textbf{Steane Code:}
The Steane Code \cite{steane1996multiple}, also known as the 7-qubit code, was introduced as a part of the Calderbank-Shor-Steane (CSS) \cite{calderbank1996good, steane1996error} family of quantum error-correcting codes. It represents a significant advancement in the field of quantum error correction, offering a symmetrical approach to correcting both bit-flip and phase-flip errors. The Steane Code maps one logical qubit to seven physical ones, drawing from the [7,4,3] Hamming code to correct single-bit errors and extending it to fix both bit-flip and phase-flip errors in qubits. Its CSS construction enables independent correction of these errors, streamlining the correction process and simplifying the required quantum circuits. The logical zero $\ket{0_L}$ and one $\ket{1_L}$ states in the Steane Code are defined using superpositions of the Hamming code's even and odd weight codewords, respectively. The Steane Code employs syndrome measurements for error detection, which uniquely indicate the presence and type of error without collapsing the quantum state. This code is comprised of 6 stabilizer generators: $IIIXXXX$, $IXXIIXX$, $XIXIXIX$, $IIIZZZZ$, $IZZIIZZ$ and $ZIZIZIZ$. Due to this structure, the Steane Code can independently correct X and Z errors using separate circuits. This dual capability is central to its functionality, allowing for the efficient correction of single-qubit errors in quantum systems. The logical $\ket{0}$ and $\ket{1}$ states of this code are: $\ket{0_L} = \frac{1}{\sqrt{8}} (\ket{0000000} + \ket{1010101} + \ket{0110011} +\ket{1100110} + \ket{0001111} + \ket{1011010} + \ket{0111100} + \ket{1101001})$ and $\ket{1_L} = X_L \ket{0_L}$; where $X_L = XXXXXXX$. Fig. \ref{fig:steane_code} depicts the error-detecting circuit of the 7-qubit Steane code.

\begin{figure}
    \centering
    \includegraphics[width=0.75\linewidth]{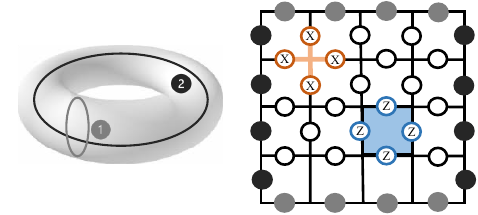}
    \caption{\textbf{Representation of a toric code.} 
    % The torus on the left shows two rings. Its corresponding lattice on the right is arranged to reflect these rings, with top-bottom and left-right lattice qubits being identical. The X-stabilizer (vertex operator), marked in orange, acts on four `X' labeled qubits to detect Z errors. Conversely, the Z-stabilizer (plaquette operator), highlighted in blue, operates on four `Z' labeled qubits, enabling the detection of X errors on any of these qubits.
    }
    \label{fig:toric_code}
    \vspace{-10pt}
\end{figure}

\textbf{Toric Code:}
The Toric Code \cite{kitaev1997quantum} represents a novel approach to quantum error correction by leveraging topological properties. Its use of a two-dimensional lattice structure is unique, making it resistant to a broad class of local errors. In the Toric Code, qubits are placed on the edges of a two-dimensional lattice on a torus. The topology of the lattice defines logical qubits and error correction is achieved by measuring vertex and plaquette operators. These operators are defined as: $\text{Vertex operators: } O_v = \prod_{i \in v} X_i \text{ and Plaquette operators: } O_p = \prod_{i \in p} Z_i$. In these expressions, $X_i$ and $Z_i$ represent Pauli X and Z operators applied to the qubits. The products are taken over qubits around a vertex $v$ for vertex operators, and around a plaquette $p$ for plaquette operators. Errors in the Toric Code manifest as excitations in this lattice structure. Bit-flip errors disrupt vertex operators, while phase-flip errors disrupt plaquette operators. The code detects errors by checking for these excitations. The topological nature of the code allows for the identification and correction of errors based on the collective state of multiple qubits, rather than individual qubit states. Fig. \ref{fig:toric_code} represents the toric code and the acting of the X and Z stabilizers on its unique lattice structure.

\begin{figure}
    \centering
    \includegraphics[width=0.75\linewidth]{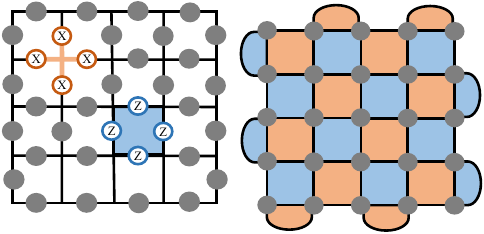}
    \caption{\textbf{Layouts of unrotated and rotated surface codes.} 
    % The lattice on the left shows an unrotated surface code which is similar to a toric code except that it is planar and is not modeled around a torus. The lattice on the right shows a rotated surface code where alternating surfaces form X and Z stabilizers.
    }
    \label{fig:surface_code}
    \vspace{-10pt}
\end{figure}

\textbf{Surface Code:}
Surface Codes \cite{dennis2002topological} are a class of quantum error-correcting codes that extend the principles of the Toric Code to a planar geometry, meaning it is not modeled around a torus. They are known for their high threshold error rate, making them a promising candidate for scalable quantum computation. Surface codes, similar to toric codes in stabilizers and error detection, have two main layouts: unrotated \cite{fowler2012surface} and rotated \cite{tomita2014low}. Unrotated surface codes use a square lattice, with data qubits on square edges, Z-stabilizers on surfaces, and X-stabilizers at vertices. Rotated surface codes feature a tilted lattice, with X and Z stabilizers alternating in a chessboard pattern. The difference in lattice arrangement impacts their error-correction efficiency, with rotated codes generally preferred for long-distance QEC due to a higher error threshold and simpler implementation. Error detection in Surface Codes is accomplished through the measurement of these stabilizer operators. A non-trivial measurement outcome (eigenvalue -1) indicates an error. The planar nature of Surface Codes facilitates the implementation of fault-tolerant gate operations and scalable quantum computing architectures. Fig. \ref{fig:surface_code} shows unrotated and rotated layouts of a surface code.

\begin{figure}
    \centering
    \includegraphics[width=0.75\linewidth]{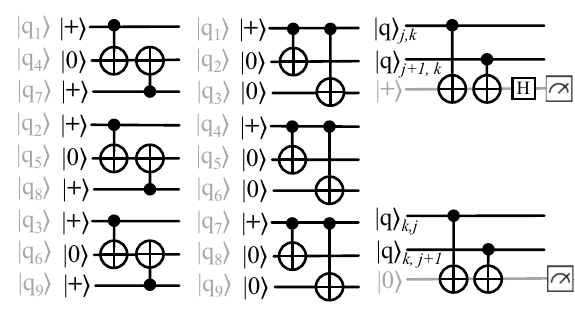}
    \caption{\textbf{Encoding and measurement in Bacon-Shor code.} 
    % The first two circuits show the encoding of logical $\ket{0}$ and $\ket{+}$ respectively for a $3 \times 3$ representation. he last two circuits show the measurement of operators $X_{j,k}X_{j+1,k}$ and $Z_{k,j}Z_{k,j+1}$ respectively with one ancilla qubits. 
    }
    \label{fig:bacon_shor_code}
    \vspace{-10pt}
\end{figure}

\textbf{Bacon-Shor Code:}
The Bacon-Shor Code \cite{bacon2006operator, shor1995scheme} is a type of quantum error-correcting code that simplifies error correction in quantum computing. It is a subclass of the more general family of Shor codes. The Bacon-Shor codes are stabilized over a square lattice, where the lattice dimensions dictate the X and Z error correction properties. The size of the lattice determines the total number of errors the code can correct. Such a code is defined as an $m_1 \times m_2$ lattice of qubits and is symmetric when $m_1 = m_2$. The X and Z stabilizers act on all qubits present on adjacent columns and rows respectively. If $O_{i,j}$ is an operator acting on the qubit at the position $(i,j)$, which $i \in \{0, 1, \dots , m_1 - 1\}$ and $j \in \{0, 1, \dots , m_2 - 1\}$. Tthe code stabilizer group is given by: $S = \langle X_{i,*}X_{i+1,*}, Z_{*,j}Z_{*,j+1} \rangle$, with generators expressed as a product of adjacent neighbor 2-qubit operators: $X_{i,*}X_{i+1,*} = \bigotimes_{k=0}^{m_2-1} X_{i,k}X_{i+1,k}$ and $Z_{*,j}Z_{*,j+1} = \bigotimes_{k=0}^{m_1-1} Z_{k,j}Z_{k,j+1}$. Syndrome extraction is done by measuring these operators, which are local and are on fewer qubits. The shortest Bacon-Shor code is $[[9,1,3,3]]$ with four stabilizer generators: $XXXXXXIII$, $IIIXXXXXX$, $ZZIZZIZZI$ and $IZZIZZIZZ$. Fig. \ref{fig:bacon_shor_code} depicts the encoding and stabilizer measurement in Bacon-Shor code \cite{suchara2013estimating, devitt2013quantum}.

\begin{figure}
    \centering
    \includegraphics[width=0.75\linewidth]{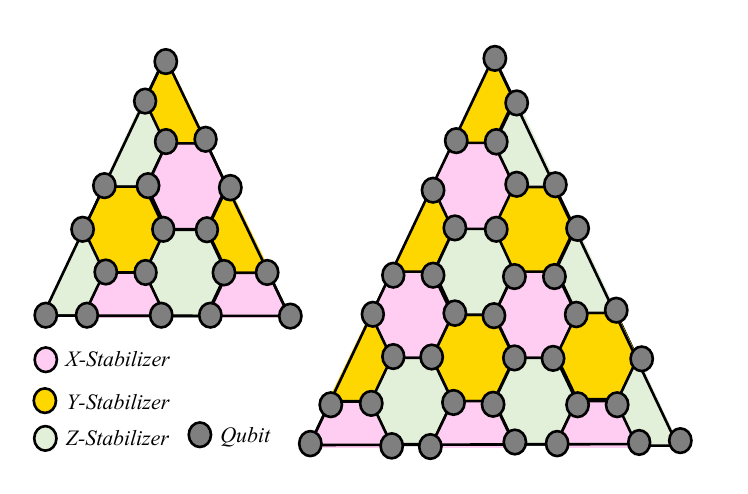}
    \caption{\textbf{Illustration of two distinct iterations of color codes.}
    % Though intrinsically three-dimensional, color codes are rendered here within a two - dimensional triangular lattice format encompassing hexagonal surfaces for simplified visual interpretation. The logical qubits, integral components of the logical state, are marked as gray-blobs. Each hexagonal surface is associated with a stabilizer, a key element for error detection and correction. Z-stabilizers are denoted by pink - hexagons, while the green - hexagons correspond to X-stabilizers. Color codes, distinguishing themselves from surface codes, incorporate an additional type of stabilizer, the Y-stabilizer, depicted by the yellow - hexagons.
    }
    \label{fig:color_code}
    \vspace{-10pt}
\end{figure}

\textbf{3D Color Code:}
The quantum color code \cite{bombin2006topological}, another form of topological QECC, takes a unique approach to error detection and correction. It extends the principles of surface codes by encoding logical qubits into a three-dimensional structure, often visualized as a two-dimensional lattice with `colors' for ease of interpretation \cite{kubica2015unfolding, chamberland2020triangular}. In our discussion, we simplify color codes into a 2D triangular lattice with hexagons, where each vertex represents a logical qubit and each hexagon a stabilizer (X, Z, or the unique Y-type for simultaneous bit and phase flips). Color codes stand out for their ability to correct a wider array of errors, including complex correlated errors, beyond the single-error correction of surface codes. Fig. \ref{fig:color_code} illustrates this with two color code instances on a 2D lattice, showing logical qubits as gray blobs and stabilizers as hexagons: pink for Z, green for X, and yellow for Y, emphasizing the role of Y-stabilizers in their advanced error-correcting capabilities.
3D Color Codes offer superior error correction by leveraging an extra spatial dimension, enhancing their ability to fix single-qubit and correlated errors—a major quantum computing hurdle. Fig. \ref{fig:color_code} showcases color codes on a 2D triangular lattice, abstracting their 3D complexity. It depicts logical qubits (gray blobs) and hexagonal stabilizers: pink for Z, green for X, and yellow for Y stabilizers, illustrating the intricate stabilizer dynamics essential for effective error correction.

\begin{figure}
    \centering
    \includegraphics[width=0.85\linewidth]{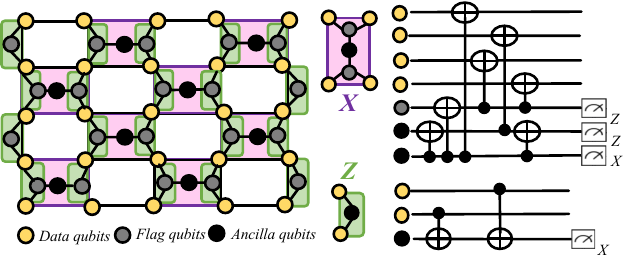}
    \caption{\textbf{Representation and stabilizer circuits of a heavy-hexagon code.} 
    % The left of the figure is the actual layout of a distance = $5$ heavy-hexagon code. The right of the figure contains two circuits to perform the X and Z parity checks respectively.
    }
    \label{fig:heavy_hexagon}
    \vspace{-10pt}
\end{figure}

\textbf{Heavy Hexagon (HH) Code:}
Heavy hexagon codes \cite{chamberland2020topological} leverage a hexagonal lattice enhanced with strategically placed 'heavy' qubits, improving error correction efficiency through increased qubit connectivity. This design boosts error tolerance, crucial for precise syndrome measurements in identifying and correcting bit-flip (X) and phase-flip (Z) errors. The mechanism ensures the integrity of quantum information by accurately pinpointing and fixing errors. This lattice stabilizer subsystem code synthesizes elements from both Bacon-Shor and surface-code frameworks, and encodes a single logical qubit into \( n = \frac{5d^2 - 2d - 1}{2} \) physical qubits where \( d \) denotes the code distance. This lattice is characterized by its sparse connectivity, typically not exceeding three connections per qubit, making it a robust candidate for systems with fixed-frequency transmon qubits that are vulnerable to frequency collision errors. On this lattice, data qubits and ancillas are strategically distributed along the hexagonal vertices and edges. A designated subset of these ancilla qubits serves as flag qubits, crucial for the detection of high-weight errors that may result from a limited number of fault events. The stabilizers tasked with the detection of X-type errors are constituted by the product of weight-two Z-type gauge operators, which collectively form the surface code stabilizers. Meanwhile, the X-type stabilizers, reminiscent of columnar operators in the Bacon-Shor code, are deduced by aggregating products of weight-four and weight-two X-type gauge operators. Fig. \ref{fig:heavy_hexagon} represents the layout and stabilizer circuits of a heavy-hexagon code.

\textbf{Bivariate Bicycle (BB) Code:}
This code \cite{bravyi2024high}, is a significant advancement in the field of quantum error correction by IBM. This innovative code belongs to the family of Abelian 2BGA codes known for their efficiency in syndrome measurement within quantum computing, particularly due to their time-optimal circuits that fit neatly into a two-layered setup. This arrangement extends the square-lattice design, optimizing it further for use with surface codes.

\begin{figure}
    \centering
    \includegraphics[width=0.6\linewidth]{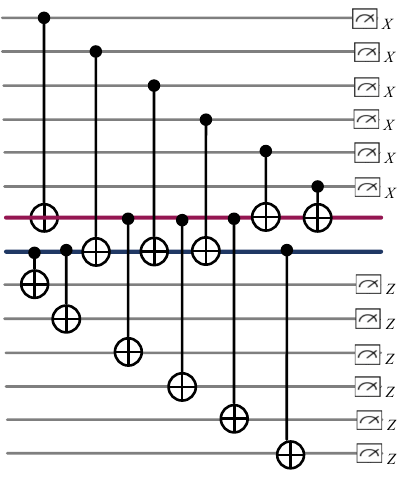}
    \caption{\textbf{Syndrome measurement circuit of a BB code.} 
   }
    \label{fig:bb}
    \vspace{-10pt}
\end{figure}

Unlike traditional 2D grid structures, the BB code's qubit connectivity graph unfolds into two planar subgraphs, each with a degree of three, offering a more complex interconnectivity. It utilizes $n$ $X$ and $Z$ check operators, each with a weight of six. The syndrome measurement circuit is shown in Fig. \ref{fig:bb}. Including the ancilla qubits necessary for operation, the BB code achieves an encoding efficiency that exceeds that of the conventional surface code. Specifically, a BB Code described by $[[n,k,d]]$ parameters necessitates $n$ ancilla qubits for encoding, resulting in an effective ancilla-adjusted encoding rate of $k/2n$. A standout among BB Codes is the Gross Code, $[[144,12,12]]$, a specific type of BB QLDPC code, which is remarkable for requiring fewer physical and ancilla qubits for syndrome extraction compared to the surface code, maintaining the same logical qubit count and distance. The term "Gross" here cleverly alludes to the code's efficiency, as "gross" traditionally means a dozen dozens, highlighting the code's compactness with an ancilla-adjusted rate of $1/24$. For comparison, the surface code with a distance of $13$ exhibits a much lower ancilla-adjusted rate of $1/338$, underscoring the BB code's superior efficiency.

\section{Benchmarking Methodology} \label{benchmarking_methods}

% This section outlines the benchmarking parameters, explains the rationale behind the selection of codes \& parameters, and discusses the methodology used for comparative analysis.

\subsection{Explanation of Benchmarking Parameters}
\circled{1}
\textbf{\textit{Qubit Overhead:}}
Qubit overhead is a fundamental parameter in assessing the efficiency of a QECC. It refers to the number of physical qubits needed to encode a logical qubit; lower overhead is preferable as it implies fewer resources for error correction. However, this must be balanced against the code's ability to correct errors, which often requires more qubits for increased distance. Therefore, analyzing the trade-off between qubit overhead and error correction capabilities is crucial in evaluating the practicality of a QECC.
\circled{2}
\textbf{\textit{Error Threshold:}}
The error threshold is a critical benchmark for the robustness of a QECC as it represents the maximum physical error rate under which a QECC can effectively correct errors. A higher threshold indicates a more fault-tolerant code, making it more desirable for practical quantum computing where error rates are non-negligible.
\circled{3}
\textbf{\textit{Error Protection:}}
Error protection capability is the measure of a QECC's ability to identify and correct a range of quantum errors, including bit flips, phase flips, and their combinations (Pauli errors). The ultimate goal is to design QECCs capable of detecting and correcting all possible quantum errors. A QECC's versatility in handling diverse error types is a significant indicator of its effectiveness and practical utility in quantum computing environments.
\circled{4}
\textbf{\textit{Decoding:}}
The availability and efficiency of decoding algorithms are essential for the practical implementation of QECCs. Decoding is the process of interpreting syndrome measurements to identify and correct errors. The more decoding algorithms a QECC is compatible with, the more flexible and adaptable it is. The performance of a QECC is heavily reliant on the efficiency of these algorithms, which should be fast and accurate to maintain the integrity of quantum information.
\circled{5}
\textbf{\textit{Transversal Gates:}}
Transversal gates play a crucial role in the implementation of quantum error correction. They facilitate operations on encoded qubits without decoding them, thereby preserving the error correction capability. The complexity and type of transversal gates that a QECC supports can significantly affect its functionality and integration with quantum circuits. Advanced techniques like lattice surgery, while complex, offer enhanced capabilities in error correction and logical operations, marking an important consideration in QECC design.
\circled{6}
\textbf{\textit{Scalability:}}
Scalability is an essential factor for the practical application of QECCs in large-scale quantum computing. It refers to a QECC's ability to maintain its error correction effectiveness as the system size increases. Scalability is crucial for the advancement of quantum computing, as it determines whether a QECC can be effectively used in more complex and realistic quantum systems.
\circled{7}
\textbf{\textit{Realization:}}
The realization of a QECC across different qubit types is a testament to its adaptability and practical relevance. The versatility of a QECC in being implemented with various qubit technologies (such as superconducting qubits, trapped-ions, etc.) is vital for its applicability in diverse quantum computing architectures. This parameter is a strong indicator of a QECC's potential for widespread adoption in the quantum computing field.
\circled{8}
\textbf{\textit{Complexity:}}
The structural and operational complexity of a QECC impacts its implementation ability and efficiency. This encompasses the intricacy of its encoding, error detection, and correction processes. A less complex QECC might be easier to implement but may offer limited error correction capabilities. Conversely, a more complex QECC might provide robust error correction but at the cost of increased resource demands and operational challenges. Balancing complexity with effectiveness and resource efficiency is key in the design and selection of QECCs for practical quantum computing.

\subsection{Criteria for Selection of Codes and Parameters}

The nine chosen quantum error correction codes illustrate the historical and conceptual evolution of QECCs. Starting from basic concepts like the quantum repetition code to more complex structures like the heavy-hexagon code, they form a kind of `family tree' that shows the development and diversification of QECCs over time. They encompass a wide array of error correction strategies, from simple error correction (quantum repetition code) to sophisticated techniques like those used in the surface and color codes. This diversity ensures a comprehensive understanding of different error correction methodologies. They vary significantly in terms of complexity and scalability. For example, the Shor code and Steane code represent more traditional approaches, whereas the toric and surface codes are geared toward large-scale quantum computing. This range allows for the assessment of QECCs across different scalability requirements. Different codes are optimized for different error models. Including a variety of codes ensures that the benchmarking study addresses a wide range of error scenarios. Some codes, like the surface code, are designed with practical implementation in mind, considering current quantum computing limitations. This inclusion allows for an assessment of QECCs not just in theory but also in their practical realization. The selection spans codes with varying qubit overheads. For instance, the quantum repetition code is simple but not efficient, whereas codes like the toric code aim to balance error correction efficacy with resource efficiency. 
%\hl{To underscore our argument and demonstrate that benchmarking is a continuous process in quantum error correction, we have also integrated IBM's newly introduced BB codes into our selection. This addition helps us evaluate their role and effectiveness within this context.} 
These nine codes collectively cover the spectrum of QECCs, from basic to advanced, from theoretical to practical. This range is crucial for a comprehensive benchmarking study. While there are other QECCs, including more might lead to redundancy without significantly adding to the understanding of QECC performance. The chosen codes are sufficiently representative of the broader field. This selection allows for a comparative analysis that can highlight the strengths and weaknesses of different approaches to quantum error correction. It enables a holistic view of the current state of QECC technology.

The eight selected parameters encompass all critical dimensions of QECC functionality - from basic resource requirements (like qubit overhead) to advanced operational aspects (like error protection and decoding). This wide-ranging coverage ensures no key aspect of QECC performance is overlooked. They strike a balance between the efficiency of a QECC (in terms of resource utilization and complexity) and its robustness (error threshold and protection capabilities). This balance is crucial for practical quantum computing where resources are limited and error rates are non-negligible. Parameters such as scalability and realization reflect the real-world challenges of implementing QECCs in various quantum computing environments. They ensure that the assessment of QECCs goes beyond theoretical effectiveness to include practical viability. Each parameter addresses a distinct aspect of QECC performance. There is minimal overlap between them, ensuring that each provides unique and valuable insight into the QECC’s capabilities. While these parameters are comprehensive based on the current understanding and needs of quantum computing, they also allow for flexibility. As quantum technology evolves, these criteria can be adapted or expanded to include new developments and requirements. Having a set of widely accepted parameters aids in standardizing the assessment of QECCs. 

\subsection{Approach for Comparative Analysis}

The analysis acknowledges that the `best' QECC depends on specific operational contexts, such as the number of qubits available, the type of errors prevalent in the system, and the practical constraints of quantum computing environments. Each selected QECC will be evaluated based on predefined criteria, including qubit overhead, error threshold, error protection, decoding, transversal gates, scalability, realization, and complexity. This structured approach ensures a comprehensive and uniform assessment of each code's capabilities and limitations. The analysis aims to highlight that while no single QECC is universally superior, each has its strengths and weaknesses. Through this structured and comprehensive approach to comparative analysis, the study will provide valuable insights into the relative merits of different QECCs. Table \ref{tab:big_table} shows the parameter analysis for all quantum error correction codes used in this study.
\section{Benchmarking Analysis} \label{benchmarking_analysis}

% This section begins by examining the performance of eight distinct QECCs against a set of eight benchmarking parameters. It then progresses to discuss the significance of benchmarking variations within existing QECCs, emphasizing the importance of such assessments.

\subsection{Comparative Benchmarking of QECCs}

\begin{table}[]
\centering
\caption{Parameter Analysis for QECCs.}
\begin{adjustbox}{angle=90}
\centering
\addtolength{\tabcolsep}{-5pt}
\setlength\extrarowheight{2.3pt}
% \small
\fontsize{9.5pt}{10.25pt}\selectfont
\begin{tabular}{c||c|c|c|c|c|c|c|c|c}
% \hline

\textbf{Parameters}                                                  & \textbf{\begin{tabular}[c]{@{}c@{}}Quantum\\ Repetition Code\end{tabular}}  & \textbf{Shor Code}                                                     & \textbf{Steane Code}                                                             & \textbf{Toric Code}                                                              & \textbf{\begin{tabular}[c]{@{}c@{}}Surface\\ Code\end{tabular}}                                & \textbf{\begin{tabular}[c]{@{}c@{}}Bacon-Shor\\ Code\end{tabular}}               & \textbf{\begin{tabular}[c]{@{}c@{}}3D Color\\ Code\end{tabular}}         & \textbf{\begin{tabular}[c]{@{}c@{}}Heavy-Hexagon\\ Code\end{tabular}}    & \textbf{\begin{tabular}[c]{@{}c@{}}BB (or Gross)\\ Code\end{tabular}}         \\ \hline \hline

\textbf{\begin{tabular}[c]{@{}c@{}}Qubit\\ Overhead\end{tabular}}    & $d$                                                                           & $9$                                                                      & $7$                                                                                & $d^2$                                                                               & $d^2$                                                                                             & $d^3$                                                                               & $\frac{(3d-1)^2}{4}$                                                                & $\frac{(5d^2 - 2d - 1)}{2}$           & $12*d$                                                   \\ 
\hline

\textbf{\begin{tabular}[c]{@{}c@{}}Error\\ Threshold\end{tabular}}   & $0.3$                                                                         & $0.3$                                                                    & $0.2$                                                                              & $0.01$                                                                              & $0.018$                                                                                           & $0.00194$                                                                         & $0.0126$                                                                    & $0.045$            &    $0.018$                                                         \\ \hline

\textbf{\begin{tabular}[c]{@{}c@{}}Error\\ Protection\end{tabular}}  & \begin{tabular}[c]{@{}c@{}}Only half of all\\ bit flip errors.\end{tabular} & \begin{tabular}[c]{@{}c@{}}Two qubits, all\\ error type\end{tabular}   & \begin{tabular}[c]{@{}c@{}}Detects errors on\\ 2 qubits, corrects 1\end{tabular} & Pauli errors                                                                 & Pauli errors                                                                               & Pauli errors                                                                 & Pauli errors                                                         & Pauli errors      & Pauli errors                                                        \\ 
\hline

\textbf{Decoding}                                                    & \begin{tabular}[c]{@{}c@{}}Automation,\\ ML\end{tabular}                    & \begin{tabular}[c]{@{}c@{}}CNOT/ H\\ Gates\end{tabular}                & \begin{tabular}[c]{@{}c@{}}CNOT/ H\\ Gates\end{tabular}                          & \begin{tabular}[c]{@{}c@{}}MWPM,\\ Tensor,\\ 10+ Algo\end{tabular}               & \begin{tabular}[c]{@{}c@{}}MWPM,\\ Tensor,\\ 10+ Algo\end{tabular}                             & \begin{tabular}[c]{@{}c@{}}Statistical\\ Mapping,\\ Check operators\end{tabular} & \begin{tabular}[c]{@{}c@{}}Restriction,\\ MaxSAT,\\ 3+ Algo\end{tabular} & \begin{tabular}[c]{@{}c@{}}MWPM, ML,\\ Neural Network,\\ 1+ Algo\end{tabular} & BP-OSD \\ \hline

\textbf{\begin{tabular}[c]{@{}c@{}}Transversal\\ Gates\end{tabular}} & None                                                                        & None                                                                   & Clifford Gates                                                                   & \begin{tabular}[c]{@{}c@{}}TPG/Lattice\\ Surgery\end{tabular}                    & \begin{tabular}[c]{@{}c@{}}Lattice\\ Surgery\end{tabular}                                      & Teleportation                                                                    & \begin{tabular}[c]{@{}c@{}}Lattice\\ Surgery\end{tabular}                & \begin{tabular}[c]{@{}c@{}}Lattice\\ Surgery\end{tabular}   & \begin{tabular}[c]{@{}c@{}}Lattice\\ Surgery\end{tabular}                  \\ 
\hline

\textbf{Scalability}                                                 & No                                                                          & No                                                                     & Yes                                                                               & Yes                                                                              & Yes                                                                                            & Yes                                                                              & Yes                                                                      & Yes               & Yes                                                            \\ 
\hline

\textbf{Realization}                                                 & \begin{tabular}[c]{@{}c@{}}Superconducting,\\ Trapped ion, etc\end{tabular} & \begin{tabular}[c]{@{}c@{}}Trapped ion,\\ Optical systems\end{tabular} & \begin{tabular}[c]{@{}c@{}}Trapped ion,\\ Rydberg atom\end{tabular}              & \begin{tabular}[c]{@{}c@{}}Superconducting,\\ Rydberg atom\\ arrays\end{tabular} & \begin{tabular}[c]{@{}c@{}}Superconducting,\\ Rydberg atom,\\ Ising anyons\end{tabular} & Trapped ion                                                                      & None                                                                     & Superconducting       & None                                                        \\ 
\hline
\textbf{Complexity}                                                  & Very Low                                                                    & Low                                                                    & Low                                                                              & Very High                                                                        & High                                                                                           & Medium                                                                           & Extremely High                                                           & Extremely High          & Very High                                                      \\ \hline \hline
\end{tabular}
\end{adjustbox}
\label{tab:big_table}
\end{table}

\begin{figure}
    \centering
    \includegraphics[width=0.9\linewidth]{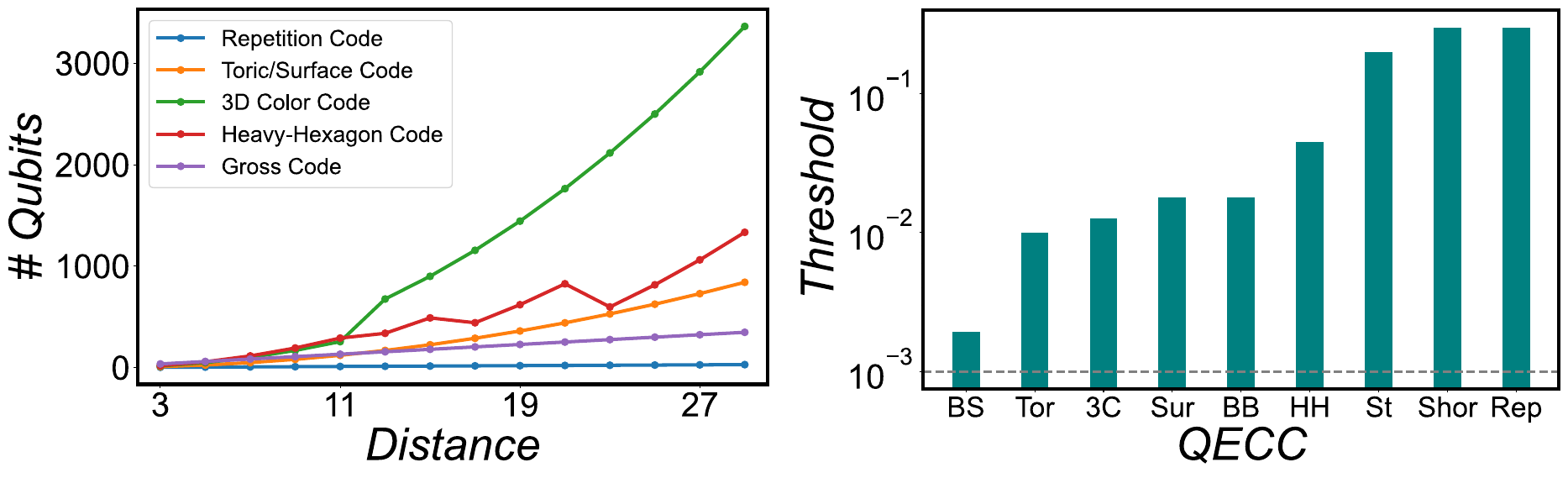}
    \caption{\textbf{Qubit overhead and threshold analysis.} 
    % The left of the figure shows the increase in the number of qubits with distance. The right of the figure shows the threshold error rate of various QECCs. 
    The grey dotted line indicates the $10^{-3}$ average error rate of an uncorrected quantum computer.
    }
    \label{fig:qecc_plots}
    \vspace{-10pt}
\end{figure}

Qubit overhead is expressed as a function of the distance variable, $d$. Specifically, the Shor and Steane codes have fixed qubit overheads of $9$ and $7$, respectively. The repetition code's overhead increases linearly with $d$. The overhead for the toric and surface codes grows at a rate proportional to $d^2$, while for the Bacon-Shor code, it escalates in proportion to $d^3$. The 3D color code and the heavy-hexagon code exhibit qubit overheads of $\frac{(3d - 1)^2}{4}$ and $\frac{5d^2 - 2d - 1}{4}$, respectively. The BB codes belong to a wider category of codes characterized by their low overhead. To quantify this overhead, it can be expressed as $n = \alpha * d^\beta$. where $\beta < 2$, indicating a subquadratic relationship. The specific values of $\alpha$ and $\beta$ vary based on the particular BB code constructions and might not be consistent across different versions. Thus, for our analysis, we will concentrate on a specific instance known as the Gross Code. The overhead for Gross Codes increases nearly linearly, at a factor of $12$ times the distance $d$. Figure \ref{fig:qecc_plots} (left) illustrates the rise in the number of qubits corresponding to increasing values of $d$. For instance, a Bacon-Shor code at a distance $11$ demands over $1000$ qubits, highlighting the significance of qubit overhead as a key benchmarking metric.

%The threshold error rate is a critical metric for quantum error correction codes (QECCs), indicating the highest level of physical errors they can effectively correct. 
The threshold error rates for various codes like the repetition, Shor, Steane, toric, surface, Bacon-Shor, 3D color, heavy-hexagon and BB codes are $3.0 \times 10^{-1}$ \cite{chatterjee2023q}, $3.0 \times 10^{-1}$ \cite{shor1995scheme}, $2.0 \times 10^{-1}$ \cite{steane1996error}, $1.0 \times 10^{-2}$ \cite{kitaev1997quantum}, $1.8 \times 10^{-2}$ \cite{chatterjee2023q, heim2016optimal}, $1.9 \times 10^{-3}$ \cite{aliferis2007subsystem}, $1.2 \times 10^{-2}$ \cite{chamberland2020triangular, delfosse2014decoding}, $4.5 \times 10^{-2}$ \cite{bhoumik2022efficient} and $1.8 \times 10^{-2}$ \cite{bravyi2024high} respectively, as depicted in Fig. \ref{fig:qecc_plots} (right). While the repetition code exhibits the highest threshold, it is important to note its limitation to only detecting bit-flip errors, rendering it impractical for broader applications. Thus, finding a balance in these characteristics is crucial for effective quantum error correction.

Different QECCs have different types and ranges of errors they can correct. The repetition code detects bit flip errors on $\lfloor (n-1)/2 \rfloor$ qubits but does not detect any phase-flip errors \cite{roffe2019quantum}. Conversely, the Shor code detects two-qubit errors or corrects an arbitrary single-qubit error \cite{shor1995scheme}. The Steane code is a distance $3$ code, detecting errors on two qubits and correcting errors on one qubit \cite{steane1996error}. More complex codes like the toric, surface, 3D color, heavy-hexagon and BB codes can detect and correct all Pauli errors. However, their effectiveness depends on the code's distance, which is directly related to the qubit overhead and the complexity of implementation. The toric code on a $d \times d$ torus is a $[[2L^2, 2, L]]$ code, encoding two qubits, while the surface code is a $[[L^2 + (L - 1)^2, 1, L]]$ code, encoding only one qubit \cite{freedman2001projective}. The Bacon-Shor code variant, $[[m_1m_2, 1, \min(m_1,m_2)]]$ code, has a distance $d = \min(m_1, m_2)$ \cite{bacon2006operator}. Unlike the surface code, the color code can suffer from unremovable hook errors due to the specifics of its syndrome extraction circuits, requiring fault-tolerant decoders to use additional flag qubits \cite{bombin2013introduction}. In contrast, the HH code is an example of the Bacon-Shor type stabilizer \cite{chamberland2020topological}.

A decoding algorithm is essential for interpreting stabilizer syndrome values to identify the type and location of errors in a quantum error correction code (QECC). These algorithms introduce additional complexity to QECCs, and the performance of these codes is highly dependent on such algorithms. The availability of multiple decoding algorithms enhances the versatility of a code. The repetition code can use classical 2D Ising model decoders \cite{toom1974nonergodic} or machine learning-based approaches \cite{convy2022machine}. The Shor and Steane codes employ CNOT and Hadamard gates for decoding \cite{nakahara2012diversities}. The toric and surface codes are supported by a variety of decoding algorithms, including minimum weight perfect-matching (MWPM) \cite{fowler2013minimum}, tensor decoders \cite{bravyi2014efficient}, union-find decoders \cite{delfosse2021almost}, brief perfect matching \cite{higgott2023improved}, renormalization group methods \cite{duclos2010fast}, Markov-chain Monte Carlo techniques \cite{hutter2014efficient}, cellular automata \cite{harrington2004analysis}, neural networks \cite{torlai2017neural}, sliding and parallel window decoders \cite{tan2023scalable}, generalized belief propagation \cite{old2023generalized}, among others. The Bacon-Shor code uses the mapping effect of noise to statistical mechanical models for decoding \cite{schmitz2020thermal}. Although its check operators are few-body, stabilizer weights scale with the number of qubits, and stabilizer expectation values are obtained by taking products of gauge-operator expectation values \cite{hastings2021fiber}. The 3D color code may utilize various decoders such as the projection decoder \cite{kubica2018abcs}, integer-program-based decoder \cite{stephens2014efficient}, restriction decoder \cite{chamberland2020triangular}, cellular-automation decoder \cite{san2023cellular}, or the MaxSAT-based decoder \cite{berent2023decoding}. The heavy-hexagon code has two decoders available primarily for itself, one using machine learning \cite{bhoumik2022efficient} and the other using a neural network \cite{hall2023artificial}, along with MWPM. The BB code can be decoded using an extended version of the BP-OSD decoder \cite{panteleev2021degenerate} to account for measurement errors.

Transversal gates facilitate operations on logically encoded qubits without the need for decoding. The choice of transversal gate contributes to the complexity of implementing a QECC. The repetition and Shor codes do not rely on transversal gates and are infrequently scalable across multiple qubits. The Steane code employs single-qubit Clifford gates, realizing the binary octahedral subgroup \cite{zeng2011transversality}. The toric, surface, 3D color, heavy-hexagon and BB codes utilize lattice surgery \cite{horsman2012surface}. In the toric code, transversal Pauli gates can be applied to non-trivial loops \cite{moussa2016transversal}. In symmetric Bacon-Shor codes, the Logical Hadamard is transversal up to a qubit permutation and can be implemented via teleportation \cite{zhou2000methodology}.

The practical implementations of QECCs reflect their applicability in real-world scenarios, and the range of qubit technologies they can be implemented on indicates their versatility. The repetition code has been realized on superconducting qubits \cite{reed2012realization, google2021exponential}, trapped-ion qubits \cite{chiaverini2004realization}, liquid-state NMR \cite{zhang2011experimental}, and NV diamond qubits \cite{waldherr2014quantum}. The Shor code has seen implementation on trapped-ion qubits \cite{nguyen2021demonstration} and in optical systems \cite{luo2021quantum}. The Steane code has been realized with trapped-ion qubits \cite{nigg2014quantum} and Rydberg atom arrays \cite{bluvstein2022quantum}. Both the toric and surface codes have been implemented using Ising anyons \cite{xu2023digital}, while the surface code has also been realized with superconducting qubits \cite{semeghini2021probing} and Rydberg atomic arrays \cite{bluvstein2022quantum}. The Bacon-Shor code's realization has been limited to trapped-ion qubits \cite{egan2020fault}. The heavy-hexagon code has been implemented in superconducting qubits \cite{sundaresan2023demonstrating}. As of now, there has been no practical realization of the 3D color and the BB codes.

The repetition and Shor codes are not scalable across multiple qubits, in contrast to the Steane, toric, surface, Bacon-Shor, 3D color, heavy-hexagon, and BB codes, which are scalable. The implementation complexity of these codes can be gauged from their circuit or lattice structures. The repetition code has a very low complexity. The Shor and Steane codes can be categorized as having low complexity. The surface code, with its square lattice structure, is considered highly complex. Similarly, the toric code, despite its resemblance in structure, is deemed highly complex due to its encoding of two qubits instead of one. Despite its low qubit overhead, the Gross Code is also considered highly complex due to its capability to encode multiple qubits simultaneously and its requirement for a two-layered setup. The Bacon-Shor code falls into the medium complexity category. Finally, the lattice structures of the 3D color and heavy-hexagon codes indicate their extremely high complexity in terms of implementation.

\begin{figure}
    \centering
    \includegraphics[width=0.9\linewidth]{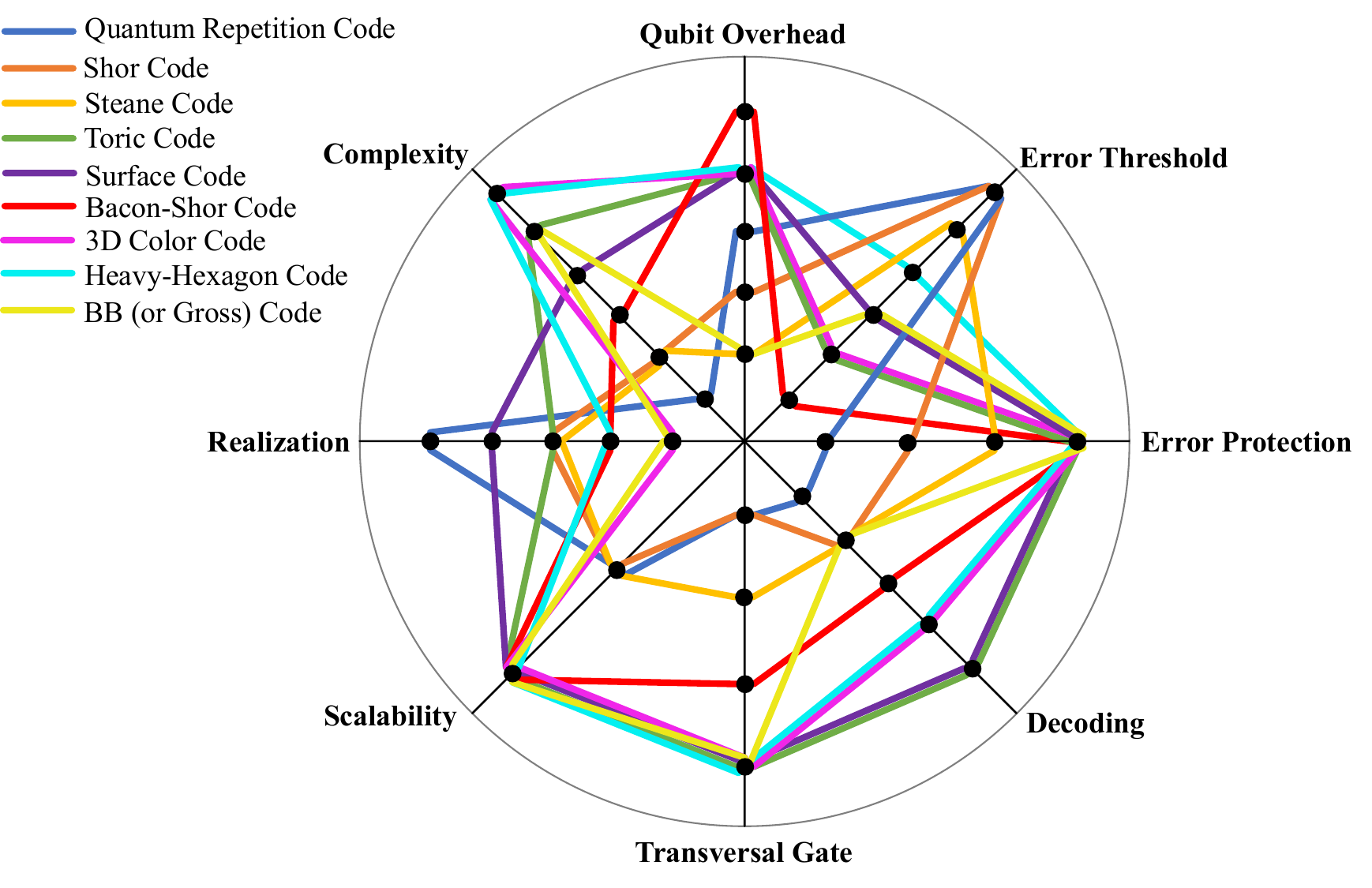}
    \caption{\textbf{Radar chart representing comparative analysis of nine QECCs across eight parameters.} 
    The parameter axes radiating from the center are as follows: Qubit Overhead = \{$7$, $9$, $d$, $d^2$, $d^3$\}; Error Threshold = \{$19^{-3}$, $0.01$, $0.018$, $0.045$, $0.2$, $0.3$\}; Error Protection = \{bit-flip, two-qubit errors, detect two-qubit errors or corrects one, all Pauli errors\}; Decoding = \{classical, $1$, $2$, $5$, $10+$\}; Transversal Gates = \{none, Clifford, teleportation, lattice surgery\}; Scalability = \{no, yes\}; Realizaton = \{none, 1, 2, 3, 6\}; and Complexity = \{very low, low, medium, high, very high, extremely high\}.
    }
    \label{fig:web_chart}
    \vspace{-15pt}
\end{figure}

Fig. \ref{fig:web_chart} benchmarks the QECCs using a radar chart which consists of eight axes, each representing a different parameter. Colored lines connecting the axes represent the QECCs. The `qubit overhead' parameter in the chart varies from constants to the highest order $d^3$. The `error threshold' spans from the lowest to the highest order value. `Error protection' encompasses types ranging from basic bit-flip to all Pauli errors. `Decoding' varies from a single classical decoder to the maximum number available for any QECC. `Transversal gates' extend from none to the most complex, lattice surgery. `Scalability' is binary, indicated by yes or no. `Realization' varies from none to the maximum number of qubit technologies on which a QECC has been implemented. The `complexity' of a QECC is on a scale of six levels, from very low to extremely high.

We observe that surface codes offer the highest form of error protection with scalability and have been realized on a considerable number of qubit technologies, with multiple decoders available. Despite having a relatively good error threshold, making them a viable option, they come at the cost of high qubit overhead, complex implementation, and the need to utilize the most complex transversal gate. Conversely, the repetition code features the least complex transversal gate and is straightforward to implement, but offers limited error protection. The choice of QECC thus becomes an art of balancing parameters and requirements.

\subsection{Benchmarking QECC Variations}

\begin{figure}
    \centering
    \includegraphics[width=0.9\linewidth]{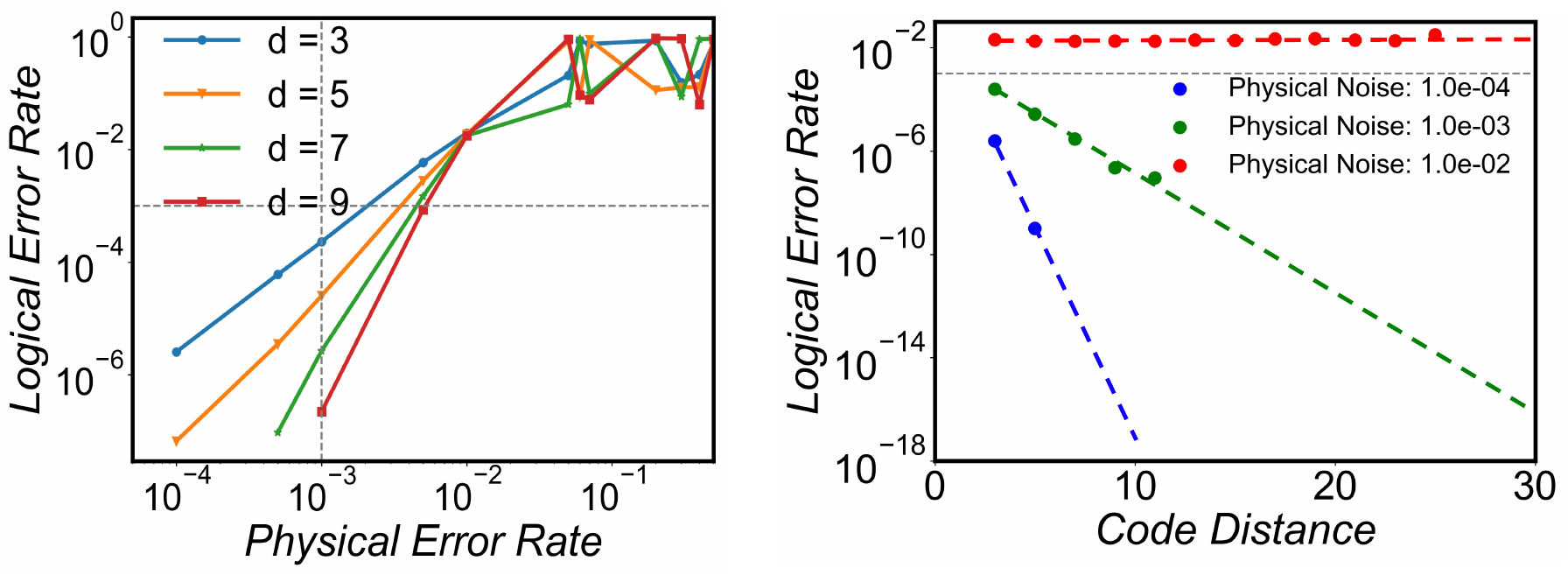}
    \caption{\textbf{Analysis of logical error rates in surface codes.} 
    % On the left, the image demonstrates how higher distance codes result in lower logical error rates for a specific physical error rate. The right side of the image illustrates the necessary increase in code distance to achieve desired logical error rate goals under different physical error rates. The grey dotted line represents the typical physical error rate of $10^{-3}$ for a quantum computer without error correction.
    }
    \label{fig:sc_plots}
    \vspace{-10pt}
\end{figure}

\begin{figure}
    \centering
    \includegraphics[width=0.9\linewidth]{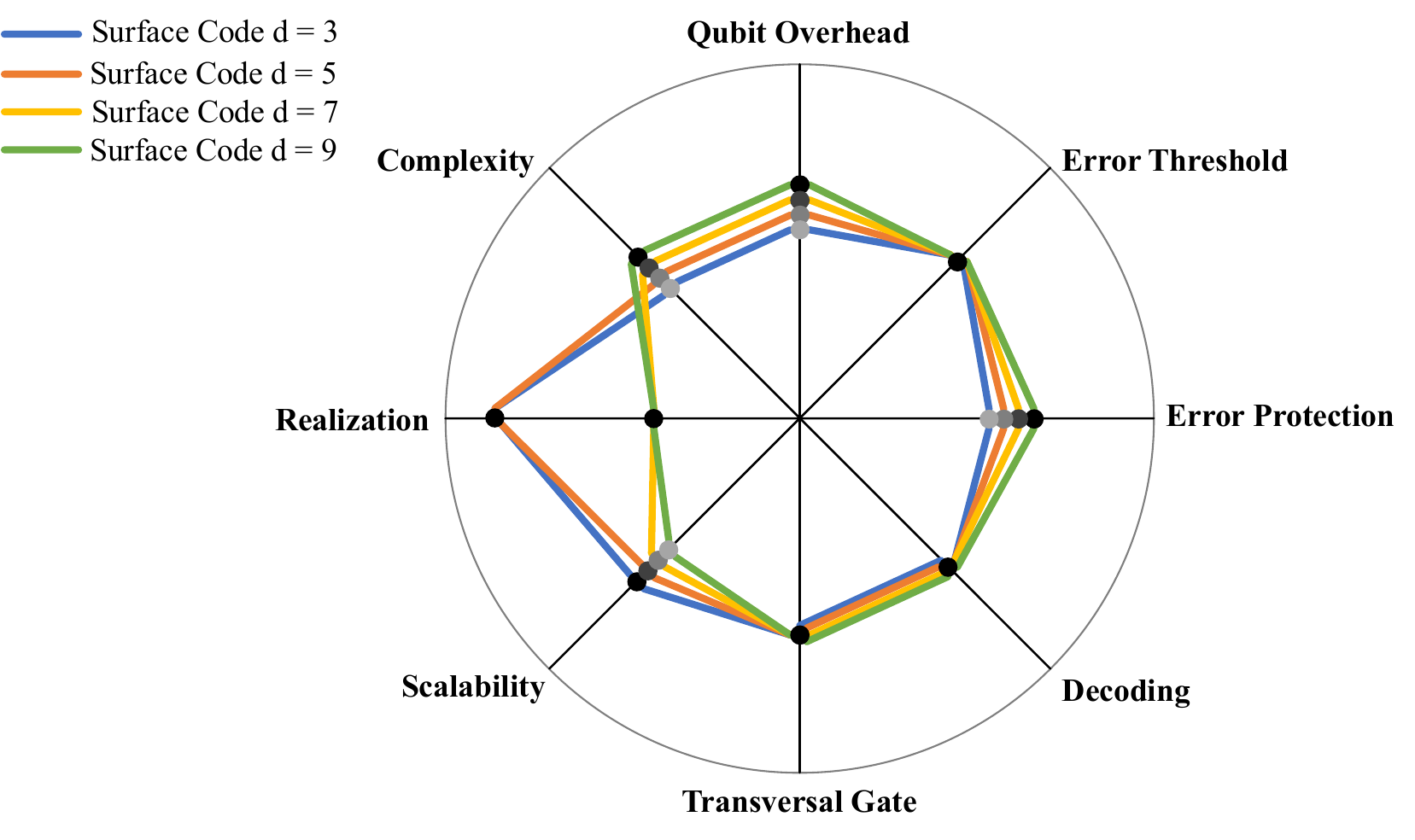}
    \caption{\textbf{Radar chart representing comparative analysis of four surface codes across eight parameters.} 
    The parameter axes radiating from the center are as follows: Qubit Overhead ($d^2$) = \{$9$, $25$, $49$, $81$\}; Error Threshold = \{$10^{-2}$\}; Error Protection = \{all Pauli errors with increasing degree\}; Decoding = \{$10+$\}; Transversal Gates = \{lattice surgery\}; Scalability = \{`yes' with increasing degree of challenges\}; Realization = \{no, yes\}; and Complexity = \{`high' with increasing degree of challenges\}.
    }
    \label{fig:web_chart_sc}
    \vspace{-10pt}
\end{figure}

It is important to understand that the radar chart presented earlier highlights only the general comparative differences among various codes. We will now focus on surface codes, which are highly applicable in real-world scenarios. For illustration, let's consider surface codes with distances of $3$, $5$, $7$, and $9$. Despite sharing common features like an identical error threshold, the same transversal gates, and equal access to decoding algorithms, these codes exhibit distinctions in several parameters. The structure of a surface code is a $d \times d$ lattice, meaning the qubit overhead increases as the distance grows. All surface codes offer comparable error protection, which is enhanced with greater distances. Figure \ref{fig:sc_plots} (left) depicts the achievable logical error rates for varying physical error rates across different surface code distances \cite{chatterjee2023q}. At a given physical error rate, a larger distance code yields a lower logical error rate. Moreover, all surface codes are scalable, but as the physical error rate rises, the necessity for greater distances becomes apparent. Although scalability is maintained with larger distances, it presents greater challenges. Figure \ref{fig:sc_plots} (right) illustrates the required distances to attain specific logical error rates for certain physical error rates \cite{chatterjee2023q}. In practical terms, only the $3$ and $5$ distance surface codes have been implemented to date \cite{bluvstein2022quantum, semeghini2021probing, krinner2022realizing}, with higher distances yet to be achieved. While the complexity of surface codes is generally considered high, it increases with distance. Figure \ref{fig:web_chart_sc} presents a radar chart comparing four different surface codes against benchmarking parameters. It is crucial to recognize that benchmarking is not only vital for evaluating different types of codes but also for assessing variations within existing codes. 
\section{Strategic Choice of a QECC} \label{strategy}

\subsection{An Universal Streamlined Approach}

\begin{figure}
    \centering \includegraphics[width=0.8\linewidth]{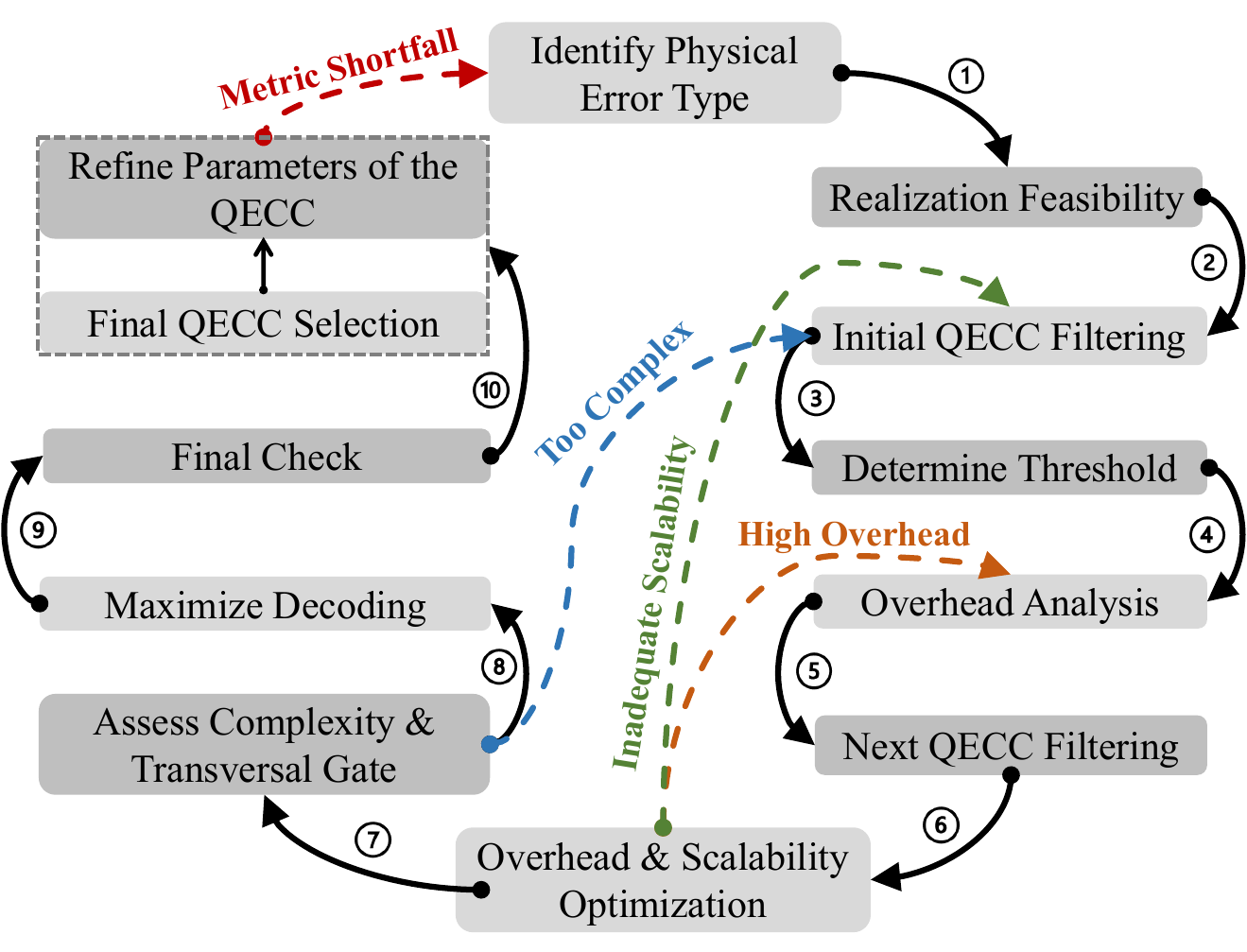}
    \caption{\textbf{A streamlined approach to QECC selection.} 
    % Illustration of the QECC selection process, from initial error type and qubit evaluation, through optimizing for overhead and scalability, to minimizing complexity and maximizing decoding options, culminating in a choice that aligns with scenario-specific requirements. Adjustments are made as needed, and feedback loops for reassessment.
    }
    \label{fig:importance}
    \vspace{-10pt}
\end{figure}

The process of selecting a code is dynamic and situational, with no fixed rules. Hence, paying attention to subtle differences in codes is as important as considering major codes. In terms of code selection, although a universal solution doesn't exist, we can establish a general systematic strategy based on typical priorities. The selection strategy for QECCs, outlined in Fig. \ref{fig:importance}, involves a multi-stage filtration process.
The strategic ordering of the multi-stage filtration process in selecting a QECC is designed to ensure a systematic and efficient narrowing down of potential codes to the one most suited for a specific scenario. This process begins with the initial evaluation of error types, qubit type, and feasibility, prioritizing the fundamental compatibility with the physical qubit system and the types of errors anticipated. This foundational step ensures that any QECC considered has the basic capability to address the primary challenges of the quantum computing system in question. Following this, the process assesses practicality through an examination of error thresholds and qubit overhead, which are critical for determining the feasibility of implementing a QECC in real-world applications. Addressing these aspects early in the selection process ensures that codes requiring unrealistic additional resources or that cannot manage expected error rates are filtered out before considering more nuanced factors. Scalability is addressed subsequently to ensure that the chosen code not only meets the immediate functional and practical requirements but also holds potential for future expansion and adaptation. This prioritization ensures that scalability does not compromise essential efficacy and practicality. Finally, the selection process considers the complexity of implementing the QECC, the ability to perform transversal gates, and the balance between complexity and decoding efficiency. These later stages refine the selection from among codes that have already passed the more fundamental tests, focusing on optimizing the balance between practical implementation and efficient operation. By following this logical sequence, the selection process ensures that the chosen QECC is not only theoretically capable of correcting anticipated errors but is also practically implementable, scalable, and efficient, balancing immediate needs with long-term potential. A last review ensures the chosen QECC meets all criteria, with adjustments made to internal parameters as needed. 
Identifying inadequacies in a QECC upon final review—such as poor error correction, high qubit overhead, scalability issues, or implementation complexities—necessitates restarting the selection process. This approach enables iterative refinement, using initial insights to more accurately meet scenario needs, accommodates evolving considerations, and adjusts to changes in resources, technology, or objectives. Thus, reassessment ensures the final QECC choice remains optimal.

\subsection{Strategic Recommendation Tool}

Our tool primarily aims to match the specific characteristics of a quantum computing scenario with suitable QECCs, recommending a list ranked from most to least favorable. This recommendation also includes the maximum feasible distance for each QECC, aligning with the understanding that longer distances typically correlate with lower logical error rates. Thus, our emphasis on maximizing distance is directly tied to achieving the lowest possible logical error rates. The framework evaluates scenarios based on five key factors: the qubit type being used, the available number of qubits for error correction within the system, the count of qubits in the circuit requiring correction, the presence of multi-qubit gates in the circuit, and the types and rates of physical errors encountered (including Pauli errors like bit-flip, phase-flip, or all-Pauli, as well as depolarizing, gate, and readout errors). Gate errors are the most weighted due to their extensive impact on the quantum system, followed by depolarizing and readout errors. Since all quantum errors can be categorized as Pauli errors, scenarios considering `all-Pauli' errors are deemed the most realistic. This flexibility aids in the versatility and characterization of QECCs. Each QECC includes a set of qubit types on which it has been implemented as one of its parameters. The `simulation' type is consistently applied across all QECCs to accommodate those not yet realized in specific scenarios, ensuring that simulated implementations are considered for every QECC. The framework aims to optimize for reduced complexity in overall operations and transversal gates while enhancing the code distance and the availability of decoding algorithms. It generates a ranked list of QECCs from most to least suitable based on these considerations, along with the maximum achievable code distance for each QECC, determined by the user’s available qubits for error correction and the qubits in the original circuit. 
% The fundamental operation of the framework is illustrated in Figure \ref{fig:magic_mirror_tool}.

% \begin{figure}
%     \centering    \includegraphics[width=0.8\linewidth]{fig/magic_mirror_tool.pdf}
%     \caption{\textbf{User inputs and outputs of the QECC recommendation tool.} 
%     % Five user inputs are depicted as entering the tool. From these inputs, the tool dynamically generates two key outputs. The creation of this tool is largely grounded in the decision-making flowchart depicted in Fig. \ref{fig:importance}.
%     }
%     \label{fig:magic_mirror_tool}
%     \vspace{-10pt}
% \end{figure}

% \usepackage{multirow}
\begin{table}[]
\centering
\caption{QECC Recommendations Based on User-Defined Scenarios.}
\begin{tabular}{cc||ccc}
\multicolumn{2}{c||}{\textbf{Parameters}}                                                                                   & \multicolumn{1}{c|}{\textbf{Scenario 1}}                                     & \multicolumn{1}{c|}{\textbf{Scenario 2}}                          & \textbf{Scenario 3} \\ \hline \hline
\multicolumn{1}{c|}{\multirow{8}{*}{\textbf{\begin{tabular}[c]{@{}c@{}}User\\ Input\end{tabular}}}} & \textbf{qType}       & supercond.                                                                   & rydberg                                                           & simulation          \\
\multicolumn{1}{c|}{}                                                                               & \textbf{maxQAvail}   & 100                                                                          & 600                                                               & 1500                \\
\multicolumn{1}{c|}{}                                                                               & \textbf{qOrig}       & 1                                                                            & 2                                                                 & 5                   \\
\multicolumn{1}{c|}{}                                                                               & \textbf{multiQGate?} & no                                                                           & yes                                                               & no                  \\
\multicolumn{1}{c|}{}                                                                               & \textbf{errType}     & bit-flip                                                                     & all-pauli                                                         & phase-flip           \\
\multicolumn{1}{c|}{}                                                                               & \textbf{depErr}      & 1E -04                                                                       & 1E -04                                                            & 1E -03              \\
\multicolumn{1}{c|}{}                                                                               & \textbf{gateErr}     & 1E -03                                                                       & 1E -03                                                            & 1E -03              \\
\multicolumn{1}{c|}{}                                                                               & \textbf{readErr}     & 1E -02                                                                       & 1E -02                                                            & 1E -01              \\ \hline \hline
\multicolumn{2}{c||}{\textbf{Recommendations}}                                                                                       & \begin{tabular}[c]{@{}c@{}}QRep, 100;\\ Surface, 10;\\ HeavyH, 6\end{tabular} & \begin{tabular}[c]{@{}c@{}}Steane, NA;\\ Surface, 17\end{tabular} & \begin{tabular}[c]{@{}c@{}}Surface, 10;\\ Gross, 10;\\ HeavyH, 7\end{tabular}          \\ \hline \hline
\end{tabular}
\label{tab:magic_mirror_tool_table}
\vspace{-10pt}
\end{table}

The framework is designed to seamlessly integrate additional QECCs and to effortlessly update existing parameters. Whether a QECC is realized on a new qubit type previously unaccounted for, or if there's an improvement in its error threshold, these changes can be rapidly incorporated. This adaptability ensures the framework remains current with advancements in the field. Moreover, the framework has an inbuilt `debug\_mode' which, when activated, allows users to observe the filtration process of QECCs step by step until only the final recommendations are left. Notably, the results are obtained almost instantaneously, highlighting the framework's efficiency and speed. Table \ref{tab:magic_mirror_tool_table} illustrates the user inputs for three scenarios and the corresponding recommendations generated by the framework.

\section{Conclusion} \label{conclusion}

This study presents a detailed benchmarking framework for Quantum Error Correcting Codes (QECCs), conducting a comparative analysis of nine QECCs across eight essential parameters. Our analysis reveals the significant impact of these parameters on QECC performance under various conditions and the trade-offs involved in selecting a QECC, highlighting the nuanced decision-making required. The study underscores the importance of continuous benchmarking in quantum computing's evolving landscape, where the choice of QECC is context-dependent. Furthermore, we propose a structured approach for QECC selection tailored to specific needs, offering a flexible and adaptable tool that generates a prioritized list of QECCs based on scenario characteristics and their maximum achievable distance. %This tool is designed to be flexible and adaptable, accommodating new QECCs and parameter adjustments, ensuring its relevance amid the rapid progression of quantum computing.
\footnote{GitHub Repository Link: \href{https://github.com/Avimita-amc8313/Magic-Mirror-on-the-Wall-How-to-Benchmark-Quantum-Error-Correction-Codes-Overall-}{Magic Mirror}}
\section*{Acknowledgment}

The work is supported in parts by the National Science Foundation (NSF) (CNS-1722557, CCF-1718474, OIA-2040667, DGE-1723687, \& DGE-1821766) and gifts from Intel. We extend our sincere gratitude to Dr. Anupam Chattopadhyay for his valuable time, wonderful ideas, and insightful suggestions that greatly contributed to this work.

\bibliographystyle{IEEEtran}
\bibliography{references}

\end{document}